\documentclass[vecphys]{svmult}

\usepackage{makeidx}         % allows index generation
\usepackage{graphicx}        % standard LaTeX graphics tool
\usepackage{multicol}        % used for the two-column index
\usepackage[bottom]{footmisc}% places footnotes at page bottom
\makeindex             % used for the subject index
\begin{document}

\title*{First Results From the Large Binocular Telescope: Deep Photometry of New dSphs}
\titlerunning{LBT Photometry of New dSphs}
\author{Matthew G.\ Coleman$^1$ \and Jelte de Jong$^1$}
\authorrunning{Coleman \& de Jong}
\institute{$^1$Max-Planck-Institut f\"{u}r Astronomie, K\"{o}nigstuhl 17, D-69117 Heidelberg, Germany
\texttt{coleman@mpia-hd.mpg.de}, \texttt{dejong@mpia-hd.mpg.de}}
\maketitle

\section{Introduction}
At the end of last century, nine dwarf spheroidal (dSph) galaxies were known to orbit the Milky Way.  This number has doubled in the past three years, with the majority of new discoveries achieved with the Sloan Digital Sky Survey.  In this contribution we present the first deep photometry for two recently discovered Local Group dwarf galaxies.  The Hercules dSph was one of five new Milky Way satellites announced last year \cite{belok07}, and Leo T (a transition dwarf) is the faintest known system with recent star formation \cite{irwin07}.  Our aim was to derive (i) an accurate structural map of these systems; and, (ii) star formation and chemical enrichment histories for both objects.  Satellite systems are known to experience tidal disruption due to the Galactic gravitational field, however there are many factors (such as the influence of the satellite halo) whose influence on this processs are not understood.  A structural map can reveal at what level the system has been distorted.  These are the first scientific results obtained with the Large Binocular Telescope.

\section{Photometry From the Large Binocular Telescope}

The Large Binocular Telescope (LBT) is located on Mount Graham in Arizona, and consists of two 8.4 metre mirrors on a common mount \cite{hill06}.  Our data were obtained as part of the LBT Science Demonstration Time during which a single mirror of the LBT was fitted with the blue channel of the Large Binocular Camera (LBC; \cite{rag06}, \cite{gial07}).  The LBC is a wide-field imager which provides a $23' \times 23'$ field of view, sampled at $0.23$ arcsec/pixel over four chips of $2048 \times 4608$ pixels.  The observations of the Hercules system consisted of 30 min in $B$, 20 min in $V$ and 25 min in $r$.  We give an expanded description of the data reduction and photometry techniques in our associated publication \cite{coleman07}.  In summary, we obtained photometry for approximately $5 \times 10^4$ sources over a $23' \times 23'$ field of view to a limiting magnitude of $V \sim 25.5$ ($1.5$ magnitudes below the Hercules main sequence turnoff).  Similarly, the Leo T observations consisted of 20 min in both the $g$ and $r$ filters, allowing a complete structural map of this system to a limiting magnitude of $g \sim 25.5$.

\section{The Elongated Hercules dSph}

The colour-magnitude diagram (CMD) of the Hercules system is shown in Fig.\ 1 (left panel).  In this diagram, we are using the $c_1$ `colour', which is a combination of photometry in $B$, $V$ and $r$ designed specifically for the Hercules system.  Essentially, if we plot a colour-colour-magnitude diagram in three dimensions (that is, $(B-V)$ vs $(V-r)$ vs $V$), the $c_1$ colour represents a ccompression of this dataset onto a two-dimensional plane which maximises the spread in colour of the Hercules stars.  This enhances the contrast in the CMD between the Hercules stars and those of the field region, and therefore allows a CMD-selection which is more effective than a simple two-filter (for example, $(B-V)$ vs $V$ space) CMD mask.

\begin{figure}
\centering
\includegraphics[height=6cm]{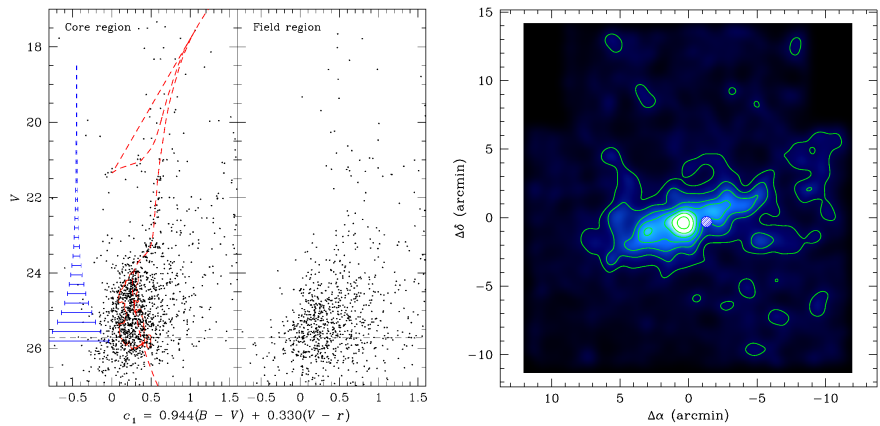}
\caption{{\em Left panel:} The Hercules and field CMDs, shown in the $c_1$ colour combining the three-filter photometry.  The dashed line is an isochrones representing the Hercules stellar population and the red contour outlines our CMD selection region.  {\em Right panel:} Structural contour diagram from the resulting CMD-cleaned dataset.  The contours represent stellar densities $1.5\sigma, 3.0\sigma,\dots,10.5\sigma$ above the background.  Both figures appear in our publication, \cite{coleman07}. \label{fig1}}
\end{figure}

CMD selection was achieved using the method described by \cite{grillmair95}, in which a `signal' map of Hercules stars compared to the field contamination is derived over the CMD.  Stars were selected in the CMD region outlined by the red contour in Fig.\ 1, we convolved them with a Gaussian of radius $0.6'$ to produce the stellar surface density contour diagram shown in Fig.\ 1 (right panel).  We have found Hercules to be highly elongated.  The ellipticity of this system ($e = 0.65$) gives a major-to-minor axis ratio of $3:1$, which is significantly greater than the $\le 2:1$ values measured for other dSphs (excluding the highly disrupted Sagittarius system).

Three scenarios suggest themselves as explanations for this unusual structure.  The first is that Hercules is cigar-shaped.  This would make it, by some margin, the most flattened of the Milky Way dSphs known, and it is unclear what this would mean for the initial formation of this system.  Also, every other system which has been observed with this level of flattening is a bright galaxy with a rotating disk.  Thus, as a second secnario, Hercules may be rotating, however this is inconsistent with our understanding of dSphs, and is not seen in a recent kinematic survey of Hercules \cite{simon07}.

The third scenario is that Hercules has been tidally distorted by the gravitational field of the Milky Way, an effect seen in other systems.  However, Hercules is relatively distant ($132 \pm 12$ kpc; \cite{coleman07}), hence tidal distortion would require this system to be on an extreme orbit.  We estimate \cite{coleman07} that a pericentric passage of $R_{\mbox{\scriptsize peri}} \sim 8$ kpc is required to have induced tidal distortion.  Hercules is not yet at apogalacticon (\cite{simon07}: Hercules is moving at 144.6 km s$^{-1}$ {\em away} from the Milky Way), hence our tidal distortion scenario requires an $e > 0.9$ orbit for Hercules.  The dSphs with known proper motions all have orbital eccentricities are all less than 0.7 (the contributions of Slawomir Piatek and Carlton Pryor).  Therefore, although we favour the tidal distortion scenario, it does suggest that Hercules is on an extreme orbit.

\section{The SFH of the Leo T Dwarf Galaxy}

\begin{figure}
\centering
\includegraphics[height=6cm]{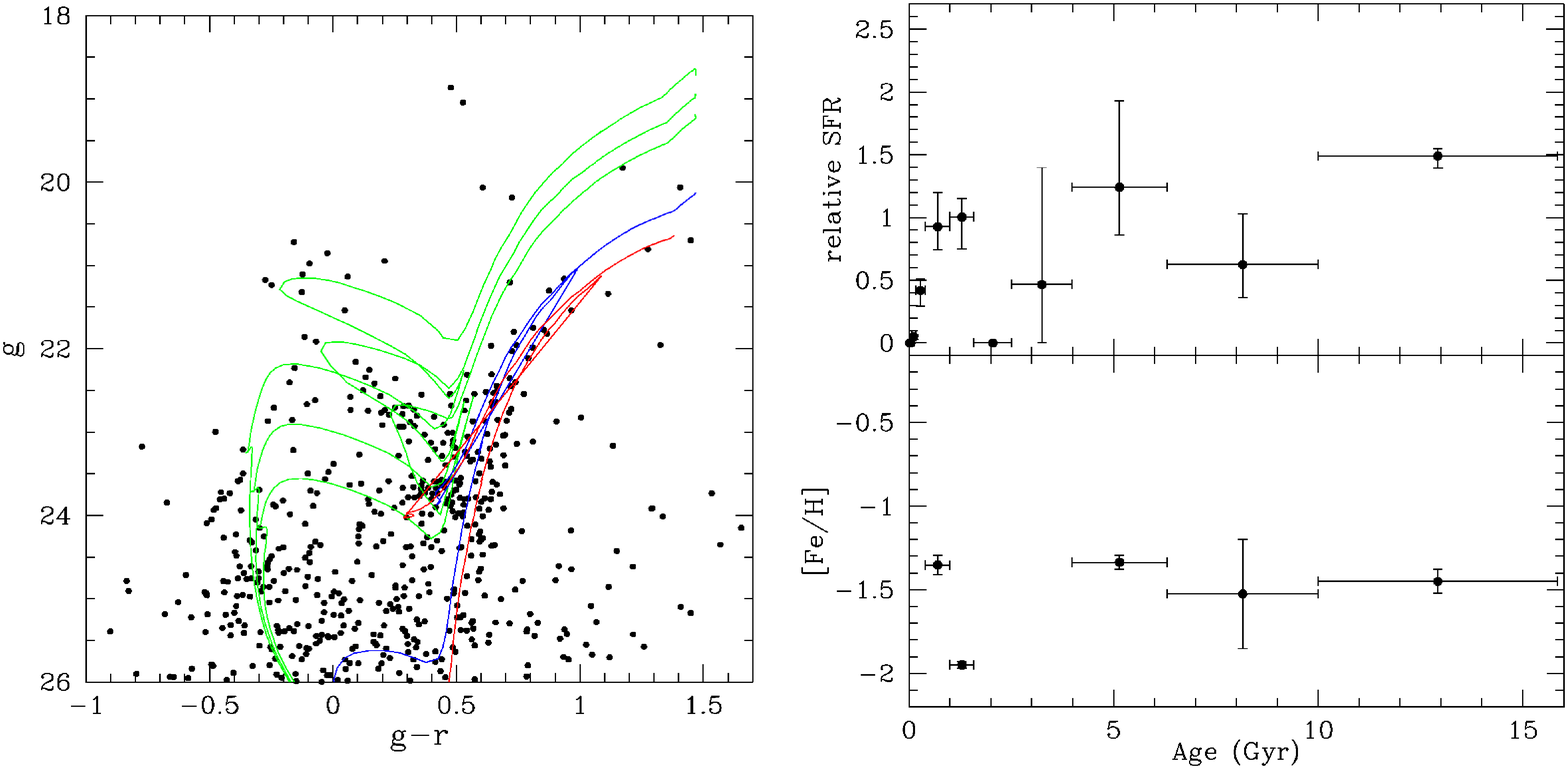}
\caption{{\em Left panel:} The Leo T CMD.  {\em Right panel:} The star formation and [Fe/H] abundance histories of Leo T.   \label{fig2}}
\end{figure}

In the left panel of Fig.\ \ref{fig2} we present the CMD of Leo T. Shown are all stars within a $1.4'$ radius from the centre of this very distant object ($\sim$420 kpc).  The previously observed \cite{irwin07} very young ($<$1 Gyr) and a much older ($>$5 Gyr) stellar populations are confirmed by our deeper photometry.  Several isochrones from \cite{girardi04} are overlayed: the green isochrones are 400 Myr, 650 Myr, and 1 Gyr isochrones with [Fe/H] $=-1.7$.  These isochrones fit the young main-sequence stars bluewards of $g-r = 0.0$ and the helium-burning blue loop stars between ($g-r,g$) of $(0.5,23.5)$ and $(-0.3,21)$.  In blue and red the 5 Gyr and 12 Gyr for [Fe/H] $=-1.7$ are shown, respectively.  Both follow the red giant branch and fit the short horizontal branch or red clump at $(g-r,g)=(0.4,23.8)$.

To study the star formation history and metallicity evolution of Leo T in more detail, the CMD-fitting software MATCH \cite{dolphin02} was used to fit the LBT photometry (see de Jong et al.\ in prep for a description of this package as applied to SDSS-filter photometry).  Stars at an appropriate distance from the centre of Leo T were used to construct a control field CMD, which was used to fit the field star contamination.  The resulting star formation rate and metallicity as function of time are plotted in the right panel of Fig.\ \ref{fig2}.  As already implied by the overlayed isochrones, an exact age is not found for the older stars, but rather continuous star formation starting in the oldest age bin and continuing until roughly 5 Gyr ago.  The young stars seem to have formed in a burst starting slightly more than 1 Gyr and ending a few hundred Myr ago.  Remarkably, we find a uniform metallicity of [Fe/H] $\simeq -1.5$ during the early star forming phase.  For the young stars we get a similar metallicity.  Because of the sparseness of this system, care should be taken not to over-interpret these early results.

\printindex
\end{document}